\documentclass[a4paper,12pt, epsfig]{article}
\pdfoutput=1
\usepackage{epsfig}
\usepackage{epstopdf}
\usepackage{graphicx}
\usepackage{ifthen}
\usepackage{feynmp}
\usepackage{amsmath}
\usepackage[psamsfonts]{amssymb}
\usepackage{euscript}

\usepackage{latexsym}
\usepackage[arrow,matrix,curve]{xy}

\newskip\humongous \humongous=0pt plus 1000pt minus 1000pt

\newif\ifdtup

\def\,{\hspace{-.1cm}}
\def\hsp{,\hspace{.7cm}}

\def\fc#1#2 {\frac{n}{q}#1\frac{n}{q}#2}

\renewcommand{\cos}{\textrm{cos}}
\renewcommand{\sin}{\textrm{sin}}

\renewcommand{\sinh}{\textrm{sinh}}
\renewcommand{\cosh}{\textrm{cosh}}
\renewcommand{\tanh}{\textrm{tanh}}
\newcommand{\sech}{\textrm{sech}}
\newcommand{\csch}{\textrm{csch}}

\def\exp#1{\hbox{\rm exp}\left(#1\right)}

\renewcommand{\theequation}{\arabic{section}.\arabic{equation}}
\renewcommand{\(}{\begin{equation}}
\renewcommand{\)}{end{equation} \vspace{-.05in}\linebreak}

\newcounter{saveeqn}
\newcounter{savealpheqn}

\newcommand{\alpheqn}{\setcounter{saveeqn}{\value{equation}}%
  \stepcounter{saveeqn}\setcounter{equation}{0}%
  \renewcommand{\theequation}{\mbox{\arabic{section}.\arabic{saveeqn}
\alph{equation}}}
  \renewcommand{\)}{\end{equation}}}
\def\part#1{\frac{\partial}{\partial{#1}}}%
\def\group#1{\refstepcounter{equation}\setcounter{saveeqn}
 {\value{equation}}%
  \label{#1}\setcounter{equation}{0}%
\renewcommand{\theequation}{\mbox{\arabic{section}.\arabic{saveeqn}
\alph{equation}}}
  \renewcommand{\)}{\end{equation}}}
\newcommand{\reseteqn}{\setcounter{equation}{\value{saveeqn}}%
  \renewcommand{\theequation}{\arabic{section}.\arabic{equation}}%
  \renewcommand{\)}{\end{equation}}}

\newcommand{\aalpheqn}{\setcounter{saveeqn}{\value{equation}}%
  \stepcounter{saveeqn}\setcounter{equation}{0}%
  \renewcommand{\theequation}{\mbox{
        \Alph{subsection}.\arabic{saveeqn}\alph{equation}}}
   \renewcommand{\)}{\end{equation}}}
\newcommand{\areseteqn}{\setcounter{equation}{\value{saveeqn}}%
  \renewcommand{\theequation}{\Alph{subsection}.\arabic{equation}}%
  \renewcommand{\)}{\end{equation}}}

\renewcommand{\thefootnote}{\alph{footnote}}
\renewcommand{\(}{\begin{equation}}
\renewcommand{\)}{\end{equation}}
\newcommand{\ba}{\begin{eqnarray}}
\newcommand{\ea}{\end{eqnarray}}
\newcommand{\cbp}{\mathop{\vtop{\ialign{##\crcr
   $\hfil\displaystyle{}\hfil$\crcr\noalign{\kern-13pt\nointerlineskip}
   \BIG{)}\hskip0pt\crcr\noalign{\kern3pt}}}}}
\newcommand{\pa}{\mathop{\vtop{\ialign{##\crcr

$\hfil\displaystyle{\oplus}\hfil$\crcr\noalign{\kern+1pt\nointerlineskip
}
   \hspace{.08in}$^{\alpha=0}$\hskip6pt\crcr\noalign{\kern3pt}}}}}
\renewcommand{\hsp}{,\hspace{.3in}}
\newcommand{\p}{^\prime}

\newcommand{\Z}{\ensuremath{\mathbb Z}}

\catcode`\@=11
\def\vereq#1#2{\lower3pt\vbox{\baselineskip1.5pt \lineskip1.5pt
\ialign{$\m@th#1\hfill##\hfil$\crcr#2\crcr\sim\crcr}}}
\catcode`\@=12

\renewcommand{\(}{\begin{equation}}
\renewcommand{\)}{\end{equation}}

\def\pin#1{\int \frac{d#1}{2\pi}}

\def\df{\mathcal{D}_f}

\newcommand{\beas}{\begin{eqnarray*}}
\newcommand{\eeas}{\end{eqnarray*}}

\newcommand{\bquo}{\begin{quote}}
\newcommand{\enqu}{\end{quote}}

\renewcommand{\Z}{{\mathbb Z}}

\def\ch{{\mathcal{H}}}
\def\co{{\mathcal{O}}}

\newcommand{\beq}{\begin{equation}}
\newcommand{\eeq}{\end{equation}}
\newcommand{\bea}{\begin{eqnarray}}
\newcommand{\eea}{\end{eqnarray}}

\newskip\humongous \humongous=0pt plus 1000pt minus 1000pt

\newif\ifdtup

\catcode`\@=11

\@addtoreset{equation}{section}

\def\@normalsize{\@setsize\normalsize{15pt}\xiipt\@xiipt
\abovedisplayskip 14pt plus3pt minus3pt%
\belowdisplayskip \abovedisplayskip
\abovedisplayshortskip \z@ plus3pt%
\belowdisplayshortskip 7pt plus3.5pt minus0pt}

\def\small{\@setsize\small{13.6pt}\xipt\@xipt
\abovedisplayskip 13pt plus3pt minus3pt%
\belowdisplayskip \abovedisplayskip
\abovedisplayshortskip \z@ plus3pt%
\belowdisplayshortskip 7pt plus3.5pt minus0pt
\def\@listi{\parsep 4.5pt plus 2pt minus 1pt
      \itemsep \parsep
      \topsep 9pt plus 3pt minus 3pt}}

\relax

\catcode`@=12

\catcode`\@=11

\def\section{\@startsection{section}{1}{\z@}{3.5ex plus 1ex minus  .2ex}{2.3ex plus .2ex}{\large\bf}}

\def\thesection{\arabic{section}}
\def\thesubsection{\arabic{section}.\arabic{subsection}}

\def\appendix{\setcounter{section}{0}
 \def\thesection{Appendix \Alph{section}}
 \def\thesubsection{\Alph{section}.\arabic{subsection}}
 \def\theequation{\Alph{section}.\arabic{equation}}}
\renewcommand{\theequation}{\arabic{section}.\arabic{equation}}

\begin{document}
\def\thefootnote{\fnsymbol{footnote}}
\def\thetitle{The Ground State of the Sine-Gordon Soliton}
\def\autone{Jarah Evslin}
\def\auttwo{Hengyuan Guo}
\def\affa{Institute of Modern Physics, NanChangLu 509, Lanzhou 730000, China}
\def\affb{University of the Chinese Academy of Sciences, YuQuanLu 19A, Beijing 100049, China}

\begin{center}
{\large {\bf \thetitle}}

\bigskip

\bigskip

{\large \noindent \autone{${}^{1,2}$} \footnote{jarah@impcas.ac.cn}}

\vskip.7cm

1) \affa\\
2) \affb\\

\end{center}

\begin{abstract}
\noindent
At one loop, we provide an explicit formula for the ground state of the one-soliton sector in the Sine-Gordon theory.  The state is given in the basis of eigenstates of the field operator, or equivalently as a Schrodinger wave functional.  The formula readily generalizes to other solitons in other models and as an example we also provide the ground state of the kink in the (1+1)-dimensional $\phi^4$ double well.

\end{abstract}

%
\setcounter{footnote}{0}
\renewcommand{\thefootnote}{\arabic{footnote}}

\section{Background Material}

The prototypical strong-weak duality is that between the Sine-Gordon model and the massive Thirring model \cite{colthir}.  The central role in this duality is played by the Sine-Gordon soliton, which becomes the fundamental fermion in the massive Thirring model.  The soliton is a solution of the classical equations of the motion of the classical Sine-Gordon theory.  In the quantum theory, it is so far understood only in a singular limit of a tree-level approximation~\cite{mandsol}.   But what lies between these two limits?   In this note, we will find the ground state of the one-soliton sector at one loop.

We will begin with a review of the Sine-Gordon soliton at one loop, treated in Ref.~\cite{rajaraman}.  The Sine-Gordon theory describes a scalar field $\phi(x)$ in 1+1 dimensions.  The Hamiltonian is
\beq
H=\int dx \ch(x) \hsp
\ch(x)=\frac{1}{2}:\pi(x)\pi(x):+\frac{1}{2}:\partial_x\phi(x)\partial_x\phi(x):-\frac{m^2}{\lambda}:\left(\cos(\sqrt{\lambda}\phi(x))-1\right):\label{hsq}
\eeq
where $m$ is a mass parameter, $\lambda$ is the coupling and $\pi$ is the conjugate momentum to $\phi$.  The field and its conjugate can be expanded in oscillator modes
\bea
\phi(x)&=&\pin{p}\phi_p e^{-ipx}\hsp
\phi_p=\frac{1}{\sqrt{2\omega_p}}\left(a^\dag_p+a_{-p}\right)\nonumber\\
\pi(x)&=&\pin{p} \pi_pe^{-ipx}\hsp
\pi_p=i \sqrt{\frac{\omega_p}{2}}\left(a^\dag_p-a_{-p}\right)
\label{osc}
\eea
where
\beq
\omega_p=\sqrt{m^2+p^2}\hsp
[a_p,a^\dag_q]=2\pi\delta(p-q)\hsp
[\phi_p,\pi_q]=2\pi i \delta(p+q). \label{acomm}
\eeq
The normal ordering convention is defined with respect to $a$ and $a^\dag$.

The classical equations of motion derived from (\ref{hsq}) admit the soliton solution
\beq
\phi_{cl}(x,t)=f(x)\hsp
f(x)=\frac{4}{\sqrt{\lambda}}\arctan{e^{mx}}. \label{ksol}
\eeq
It will be convenient to define a new Hamiltonian $H_K$ which is related to the original Hamiltonian by a similarity transformation
\beq
H_K=\df^{-1}H\df\hsp
\df={\rm{exp}}\left(-i\int dx f(x)\pi(x)\right) \label{df}
\eeq
where $\df$ is the translation operator which shifts the field by the soliton solution.  In classical field theory, $H_K$ describes oscillations about the soliton configuration.  As we have normal-ordered the Hamiltonian in (\ref{hsq}), the theory is finite and regularization is not necessary.  However, had we adapted a more general regulator, which would be necessary in a theory with fermions or more dimensions, then it is essential that the full, regulated $H$ be used in Eq.~(\ref{df}).

Let $|K\rangle$ be the soliton ground state and, motivated by \cite{hepp,taylor78}, define $\co$ to be any operator such that
\beq
|K\rangle=\df \co|0\rangle. \label{cdef}
\eeq
If $E$ is the minimum soliton energy, then
\beq
E|K\rangle=H|K\rangle=\df H_K\co|0\rangle.
\eeq
Left multiplying by $\df^{-1}$ one finds
\beq
H_K\co|0\rangle=E\co|0\rangle \label{prob}
\eeq
and so $\co|0\rangle$ is the lowest eigenvector of $H_K$. 

It was shown in Ref.~\cite{hengyuan} that $\co$ is equal to the identity plus quantum corrections and so (\ref{prob}) can be solved in perturbation theory.  In this paper we will work at one-loop.  Let $Q_0$  be the classical soliton energy.  Then $H_K=Q_0+\int dx \ch_{PT}$ is the free P\"oschl-Teller Hamiltonian with Hamiltonian density
\beq
\ch_{PT}= \frac{:\pi^2(x):}{2}+\frac{:\partial_x\phi(x)\partial_x\phi(x):}{2}+\left(\frac{m^2}{2}-m^2{\rm{sech}}^2\left(mx\right)\right):\phi^2(x):
 \label{hpt}
\eeq
whose classical equations of motion have constant frequency solutions $g_k(x)$ parameterized by $k$ and a bound state solution $g_B(x)$ representing the soliton Goldstone mode
\beq
g_k(x)=\frac{e^{-ikx}}{\sqrt{1+m^2/k^2}}\left(1-i\frac{m}{k}\tanh(m x)\right)\hsp
g_{B}(x)=\sqrt{\frac{m}{2}}\sech\left(mx\right)  \label{geq}
\eeq
with respective frequencies
\beq
\omega_k=\sqrt{m^2+k^2}\hsp
\omega_B=0.
\eeq
These have been normalized so that
\beq
\int dx g_{k_1} (x) g^*_{k_2}(x)=2\pi \delta(k_1-k_2)\hsp
\int dx |g_{B}(x)|^2=1 \label{comp}
\eeq
and they satisfy the reality conditions
\beq
g_k^*(x)=g_{-k}(x)\hsp g_B^*(x)=g_B(x).
\eeq

In the ground state sector it was convenient to decompose the field $\phi(x)$ into plane waves (\ref{osc}) to obtain the Heisenberg operators $a_p$.  Similarly, in the one-soliton sector it is convenient to decompose $\phi(x)$ into the constant frequency solutions (\ref{geq})
\bea
\phi(x)&=&\phi_0 g_{B}(x)+\pin{k}\phi_kg_k(x)\hsp
\pi(x)=\pi_0 g_{B}(x)+\pin{k}\pi_kg_k(x)\nonumber\\
\phi_k&=&\frac{1}{\sqrt{2\omega_k}}\left(b_k^\dag+b_{-k}\right)\hsp
\pi_k=i\sqrt{\frac{\omega_k}{2}}\left(b_k^\dag - b_{-k}\right).
\eea
Using the completeness relations (\ref{comp}) these can be inverted
\bea
\phi_0&=&\int dx \phi(x) g_B^*(x)\hsp
\pi_0=\int dx \pi(x) g_B^*(x)\hsp
\phi_k=\int dx \phi(x) g_k^*(x)\\
\pi_k&=&\int dx \pi(x) g_k^*(x)\hsp
b_k^\dag=\sqrt{\frac{\omega_k}{2}}\phi_k-\frac{i}{\sqrt{2\omega_k}}\pi_k\hsp 
b_{-k}=\sqrt{\frac{\omega_k}{2}}\phi_k+\frac{i}{\sqrt{2\omega_k}}\pi_k\nonumber
\eea
which fix the commutation relations
\beq
[\phi_p,\pi_q]=2\pi i\delta(p+q)\hsp
[\phi_0,\pi_0]=i\hsp 
[b_{k_1},b^\dag_{k_2}]=2\pi\delta(k_1-k_2).
\eeq

After a tedious calculation one finds \cite{hengyuan}
\beq
H_K=Q_1+\pin{k}\omega_k b^\dag_k b_k+\frac{\pi_0^2}{2}
\eeq
where $Q_1$ is the one-loop soliton energy.   Thus the lowest eigenstate (\ref{prob}) solves
\beq
b_k\co|0\rangle=\pi_0\co|0\rangle=0. \label{coeq}
\eeq
So finishes our review of Ref.~\cite{hengyuan}.

\section{The Sine-Gordon Soliton State}

We will work in the basis of states provided by the eigenvectors $|F\rangle$ of the field operator $\phi(x)$.  Here $F$ is a real-valued function of the space coordinate $x$.  The basis states are defined by the eigenvalue equation
\beq
\phi(x)|F\rangle=F(x)|F\rangle.
\eeq  
Any state $|\Psi\rangle$ can be expanded in terms of the basis states $F$
\beq
|\Psi\rangle=\int DF \Psi[F] |F\rangle
\eeq
where $DF$ is a measure on the space of functions $F$ and $\Psi[F]$ is the complex-valued Schrodinger wave functional $\Psi$ \cite{stuck,fried} evaluated on the function $F$.  This is the analogue of a wave function in quantum mechanics
\beq
|\psi\rangle=\int dx \psi(x)|x\rangle.
\eeq
Just as in quantum mechanics, states can be equivalently described by Dirac kets $|\Psi\rangle$ or their matrix elements, the wave functionals $\Psi[F]$ and in fact both of these objects satisfy the same operator equations.  Thus for simplicity we will work with the wave functionals and not the kets.

Before solving  Eq.~(\ref{coeq}) for the soliton state, let us remind the reader of the solution of
\beq
a_p|0\rangle=0
\eeq
which describes the ground state of the free massive scalar theory.   Working in terms of wave functionals, this is
\beq
a_p\Psi_0=0.
\eeq
Inverting (\ref{osc})
\beq
a_{-p}=\sqrt{\frac{\omega_p}{2}}\phi_p+\frac{i}{\sqrt{2\omega_p}}\pi_p.
\eeq
Let us try
\beq
\Psi_0=\exp{-\frac{1}{2}\pin{q}\phi_q\omega_q\phi_{-q}}.
\eeq
Using the commutator in (\ref{acomm}) one finds
\beq
\pi_p\Psi_0=\pin{q}(-2\pi i) \delta(p+q)\left(-\frac{1}{2}\right)2\omega_q\phi_{-q}\Psi_0=i \omega_p\phi_p\Psi_0 
\eeq
and so
\beq
a_{-p}\Psi_0=\sqrt{\frac{\omega_p}{2}}\phi_p\Psi_0+\frac{i}{\sqrt{2\omega_p}}\pi_p\Psi_0=0
\eeq
establishing that $\Psi_0$ is the ground state wave functional of the vacuum sector.

Similarly, for the soliton wave functional corresponding to $\co|0\rangle$ try
\beq
\Psi_\co=\exp{-\frac{1}{2}\pin{k}\phi_k\omega_k\phi_{-k}}.
\eeq
As $\Psi_\co$ is independent of $\phi_0$ and $\pi_0$ commutes with $\phi_k$, trivially
\beq
\pi_0 \Psi_\co.
\eeq
In other words, this state is translation-invariant and so it has zero momentum.  Following the argument above
\beq
\pi_k\Psi_\co=i \omega_k\phi_k\Psi_\co
\eeq
and so $\Psi_\co$ is annihilated by $b_{-k}$ and solves (\ref{coeq}).  Therefore we have shown that
\beq
\co|0\rangle=\int DF \Psi_\co[F] |F\rangle.
\eeq

According to the definition (\ref{cdef}), to obtain the kink state, we need to left multiply this result by
\beq
\df=\exp{-i\pi_0 f_B-i\pin{k} \pi_k f_k}\hsp
f_B=\int dx f(x) g_B(x)\hsp
f_k=\int dx f(x) g_k(x).
\eeq
Using the fundamental property of the translation operator
\beq
[\df,\phi(x)]=-f(x)\phi(x)
\eeq
and the completeness relations (\ref{comp}) one finds the action of $\df$ on $\phi_0$ and $\phi_k$
\beq
[\df,\phi_0]=-f_B\df\hsp
[\df,\phi_k]=-f_{-k}\df.
\eeq
Thus we can compute the action of $\df$ on any function of $\phi_0$ and the set of $\phi_k$, it simply translates each argument.  The soliton ground state is then
\beq
\Psi_K=\df\Psi_\co=\exp{-\frac{1}{2}\pin{k}\left(\phi_k-f_{-k}\right)\omega_k\left(\phi_{-k}-f_k\right)}.
\eeq
This is our main result.

\section{The Double Well Kink State}

We expect the above construction to apply to stationary classical solutions in a range of quantum field theories.  In this section we will show how trivially it is extended to the $\phi^4$ double well in 1+1 dimensions, first treated at one loop in Ref.~\cite{dhn2}.  This theory is described by the Hamiltonian 
\beq
H=\int dx \left[\frac{1}{2}:\pi(x)\pi(x):+\frac{1}{2}:\partial_x\phi(x)\partial_x\phi(x):+\frac{\lambda}{4} :\phi^2(x)\left(\phi(x)-2v\right)^2:\right]
\eeq
and has a classical kink solution
\beq
\phi_{cl}(x,t)=f(x)\hsp
f(x)=\frac{m}{\sqrt{2\lambda}}\left(1+{\rm{tanh}}\left(\frac{mx}{2}\right)\right). \label{kink}
\eeq
Again we define a kink Hamiltonian by the similarity transform (\ref{df}), where $f(x)$ is now given by Eq.~(\ref{kink}).  At one loop the corresponding Hamiltonian density is
\beq
\ch_{PT}= \frac{:\pi^2(x):}{2}+\frac{:\partial_x\phi(x)\partial_x\phi(x):}{2}+\left(2\beta^2-3\beta^2{\rm{sech}}^2\left(\beta x\right)\right):\phi^2(x):
\eeq
where $\beta=m/2$.  Like (\ref{hpt}) in the case of the Sine-Gordon theory, this is a reflectionless P\"oschl-Teller Hamiltonian.  However, now it is at level 2 instead of level 1, due to the factor of 3.

The difference in level in the potential means that in addition to the Goldstone mode $g_B(x)$, $H_K$ has an additional, odd, classical bound state $g_{BO}(x)$.  Overall the eigenfunctions are
\bea
g_k(x)&=&\frac{e^{-ikx}}{\sqrt{(1+\beta^2/k^2)(1+4\beta^2/k^2)}}\left(1+\frac{\beta^2}{k^2}\left(1-3\tanh^2(\beta x)\right)-3i\frac{\beta}{k}\tanh(\beta x)\right)\nonumber\\
g_B(x)&=&\frac{\sqrt{3\beta}}{2}\sech^2(\beta x)\hsp 
g_{BO}=-i\sqrt{\frac{3\beta}{2}}\tanh(\beta x)\sech(\beta x).
\eea
The frequencies, completeness relations and reality conditions are as in the Sine-Gordon case except now we also have
\beq
\omega_{BO}=\beta\sqrt{3}\hsp
\int dx |g_{BO}(x)|^2=1\hsp
 g_{BO}^*(x)=-g_{BO}(x).
\eeq

Let us decompose the field and its conjugate
\bea
\phi(x)&=&\phi_0 g_{B}(x)+\phi_{BO}g_{BO}(x)+\pin{k}\phi_kg_k(x)\nonumber\\
\pi(x)&=&\pi_0 g_{B}(x)+\pi_{BO}g_{BO}(x)+\pin{k}\pi_kg_k(x)\nonumber\\
\phi_{BO}&=&\frac{1}{\sqrt{2\omega_{BO}}}\left(b^\dag_{BO}-b_{BO}\right)\hsp
\pi_{BO}=i\sqrt{\frac{\omega_{BO}}{2}}\left(b^\dag_{BO}+b_{BO}\right).
\eea
Using the completeness relations (\ref{comp}) these can be inverted
\bea
\phi_{BO}&=&\int dx \phi(x) g_{BO}^*(x)\hsp
\pi_{BO}=\int dx \pi(x) g_{BO}^*(x)\nonumber\\
b^\dag_{BO}&=&\sqrt{\frac{\omega_{BO}}{2}}\phi_{BO}-\frac{i}{\sqrt{2\omega_k}}\pi_{BO}\hsp 
b_{BO}=-\sqrt{\frac{\omega_{BO}}{2}}\phi_k-\frac{i}{\sqrt{2\omega_{BO}}}\pi_{BO}
\eea
which fix the commutation relations
\beq
[\phi_{BO},\pi_{BO}]=-i\hsp[\phi^*_{BO},\pi_{BO}]=i\hsp
[b_{BO},b^\dag_{BO}]=1
. \label{wcomm}
\eeq
Note that $\phi_{BO}$ and $\pi_{BO}$ are anti-Hermitian 
\beq
\phi_{BO}^*=-\phi_{BO}\hsp \pi_{BO}^*=-\pi_{BO}
\eeq
which is the reason for the wrong sign in Eq.~(\ref{wcomm}).  We adopt a star instead of a dagger for Hermitian conjugation of $\phi_{BO}$ for later convenience, because the field $\phi$ in a Schrodinger wave functional is interpreted as a function and not an operator.  However strictly speaking a dagger should be used at this step, as $\phi_{BO}$ is an operator, to be replaced by a star only at the end of the calculation when we write the wave functional.  The interpretation of $\phi$ as a function, instead of an operator, in the wave functional is possible because $\pi$ does not appear, and so $\phi$ commutes with everything.  In quantum mechanics the analogous statement is that $x$ is an operator, but when the wave function is presented $x$ may be interpreted as simply a coordinate.

After some calculation one finds \cite{kink}
\beq
H_K=Q_1+\pin{k}\omega_k b^\dag_k b_k+\omega_{BO}b^\dag_{BO} b_{BO}+\frac{\pi_0^2}{2}.
\eeq
Thus the lowest eigenstate solves
\beq
b_k\co|0\rangle=b_{BO}\co|0\rangle=\pi_0\co|0\rangle=0. 
\eeq
This is identical to the Sine-Gordon case except for the $b_{BO}$ condition, which states that the quantum harmonic oscillator corresponding to this oscillation mode of the kink is in its ground state.  To solve this condition note that
\beq
\pi_{BO}\exp{-\frac{1}{2}\phi_{BO}\omega_{BO}\phi^*_{BO}}=i\phi_{BO}\omega_{BO}\exp{-\frac{1}{2}\phi_{BO}\omega_{BO}\phi^*_{BO}}
\eeq
and so
\beq
b_{BO}\exp{-\frac{1}{2}\phi_{BO}\omega_{BO}\phi^*_{BO}}=0.
\eeq
Thus the Schrodinger wave functional corresponding to $\co|0\rangle$ is
\beq
\Psi_\co=\exp{-\frac{1}{2}\phi_{BO}\omega_{BO}\phi^*_{BO}-\frac{1}{2}\pin{k}\phi_k\omega_k\phi^*_{k}}.
\eeq

As in the Sine-Gordon case, to obtain the kink ground state we need to multiply by
\beq
\df=\exp{-i\pi_0 f_b-i\pi_{BO}f_{BO}-i\pin{k} \pi_k f_k}\hsp
f_{BO}=\int dx f(x) g_{BO}(x).
\eeq
The wrong-sign canonical commutation relations in (\ref{wcomm}) lead to
\beq
[\df,\phi_{BO}]=f_{BO}\df\hsp
[\df,\phi^*_{BO}]=f^*_{BO}\df=-f_{BO}\df
\eeq
and so the kink ground state Schrodinger wave functional is
\beq
\Psi_K=\df\Psi_\co=\exp{-\frac{1}{2}(\phi_{BO}+f_{BO})\omega_{BO}(\phi^*_{BO}+f^*_{BO})-\frac{1}{2}\pin{k}\left(\phi_k-f_{-k}\right)\omega_k\left(\phi_{-k}-f_k\right)}.
\eeq

\section* {Acknowledgement}

\noindent
JE is supported by the CAS Key Research Program of Frontier Sciences grant QYZDY-SSW-SLH006 and the NSFC MianShang grants 11875296 and 11675223.   JE also thanks the Recruitment Program of High-end Foreign Experts for support.

  \end{document}